\documentstyle[12pt,amsfonts]{article}
\input epsf.tex

\newcommand{\bmat}{\left(\begin{array}}
\newcommand{\emat}{\end{array}\right)}
\def\NPB#1#2#3{Nucl. Phys. B{#1} (19#2) #3}
\def\PLB#1#2#3{Phys. Lett. B{#1} (19#2) #3}

\def\PRD#1#2#3{Phys. Rev. D{#1} (19#2) #3}
\def\PRL#1#2#3{Phys. Rev. Lett. {#1} (19#2) #3}

\def\yzero{\smash{\hbox{$y\kern-4pt\raise1pt\hbox{${}^\circ$}$}}}

\def\beq{\begin{equation}}
\def\eeq{\end{equation}}
\def\beqa{\begin{eqnarray}}
\def\eeqa{\end{eqnarray}}

\def\-{\hphantom{-}}

\def\s2{\frac{1}{2}}

\def\beq{\begin{equation}}
\def\eeq{\end{equation}}
\def\beqa{\begin{eqnarray}}
\def\eeqa{\end{eqnarray}}
\def\vev#1{\langle#1\rangle}

\def\Tr{{\rm Tr \,}}

\def\IF{\relax{\rm I\kern-.18em F}}
\def\II{\relax{\rm I\kern-.18em I}}
\def\IP{\relax{\rm I\kern-.18em P}}
\def\IC{\relax{\rm I\kern-.48em C}}
\def\IR{\relax{\rm I\kern-.18em R}}

\def\cn{{\cal N}}

\def\cp{{\cal P}}

\def\id{{\bf 1}}

\def\Dsl{\,\raise.15ex\hbox{/}\mkern-13.5mu D} 
\def\IZ{{\Bbb Z}}
\def \one{\relax{\rm 1\kern-.26em I}}
\def\id{\one}

 \def\cp#1{\relax\ifmmode {\IP\kern-2pt{}_{#1}}\else $\IP\kern-2pt{}_{#1}$\=fi}

\catcode`\@=11
\newdimen\@rotdimen
\newbox\@rotbox

\def\@vspec#1{\special{ps:#1}}
\def\@rotstart#1{\@vspec{gsave currentpoint currentpoint translate
   #1 neg exch neg exch translate}}
\def\@rotfinish{\@vspec{currentpoint grestore moveto}}
\def\@rotr#1{\@rotdimen=\ht#1\advance\@rotdimen by\dp#1%
   \hbox to\@rotdimen{\hskip\ht#1\vbox to\wd#1{\@rotstart{90 rotate}%
   \box#1\vss}\hss}\@rotfinish}
\def\@rotl#1{\@rotdimen=\ht#1\advance\@rotdimen by\dp#1%
   \hbox to\@rotdimen{\vbox to\wd#1{\vskip\wd#1\@rotstart{270 rotate}%
   \box#1\vss}\hss}\@rotfinish}%
\def\@rotu#1{\@rotdimen=\ht#1\advance\@rotdimen by\dp#1%
   \hbox to\wd#1{\hskip\wd#1\vbox to\@rotdimen{\vskip\@rotdimen
   \@rotstart{-1 dup scale}\box#1\vss}\hss}\@rotfinish}%
\def\@rotf#1{\hbox to\wd#1{\hskip\wd#1\@rotstart{-1 1 scale}%
   \box#1\hss}\@rotfinish}%
\def\rotate{\@ifnextchar[{\@rotate}{\@rotate[l]}}
\def\@rotate[#1]#2{\setbox\@rotbox=\hbox{#2}\@nameuse{@rot#1}\@rotbox}

\catcode`\@=12

\begin{document}

\makeatletter \@addtoreset{equation}{section} \makeatother
\renewcommand{\theequation}{\thesection.\arabic{equation}}
\pagestyle{empty}
\rightline{DAMTP-2001-4, HU-EP-01/02,  \tt hep-th/0101186}
\vspace*{1.0in}

\vspace{0.3cm}
\begin{center}
\LARGE{\bf  
The Dilaton Potential from \boldmath{${\cal N}= 1^*$}  \\[10mm]}
\smallskip
\large{Anamar\'{\i}a Font${}^\flat$\footnote{On leave of absence from
Departamento de F\'{\i}sica, Facultad de Ciencias,
Universidad Central de Venezuela. Work supported by a fellowship from the
Alexander von Humboldt Foundation.}, 
Matthias Klein${}^\sharp$ and  Fernando Quevedo${}^\sharp$ 
\\[2mm]}
\small{${}^\flat$ Humboldt Universit\"at zu Berlin,\\[-0.3em]
Institut f\"ur Physik, Invalidenstr. 110, 10115 Berlin, Germany\\[1mm]
${}^\sharp$ DAMTP, Wilberforce Road, Cambridge, CB3 0WA, England.\\[1mm]

}

\end{center}
\vspace{2cm}

{\small
\begin{center}
\begin{minipage}[h]{15.5cm}
Recent understanding of ${\cal N}=1^*$ supersymmetric theory
(mass deformed ${\cal N}=4$) has made it possible to 
find an exact  superpotential which encodes the properties 
of the different phases of the theory.
We consider this superpotential as an illustrative example 
for the source of a nontrivial  scalar potential for the 
string theory dilaton and study its properties.
The superpotential is characterized by the rank of the corresponding
gauge group ($N$) and integers $p,q,k$ labelling the different
massive phases of the theory. For generic values of these parameters,
we find the expected runaway behaviour of the potential to vanishing 
string coupling. But there are also supersymmetric minima at weak 
coupling stabilizing the dilaton field. An interesting property of 
this potential is that there is a proliferation of supersymmetric 
vacua in the confining phases, with the number of vacua increasing 
with $N$ and leading to a kind of staircase potential.
For a range of parameters, it is possible to obtain realistic
values for the gauge coupling. 
\end{minipage}
\end{center}
}
\newpage

\setcounter{page}{1} \pagestyle{plain}
\renewcommand{\thefootnote}{\arabic{footnote}}
\setcounter{footnote}{0}


\section{Introduction}

The main obstacles for string theory to make contact with low-energy
physics have been the breakdown of supersymmetry and the lifting of 
the large degeneracy of string vacua which manifests itself by the 
existence of several fields with flat potentials in the low-energy 
supersymmetric theory: the dilaton and moduli fields.
 The first problem has been recently
reanalysed in the light of low fundamental scale D-brane models in 
the sense that  realistic string models have been obtained with
supersymmetry explicitly broken by the presence of anti-D-branes 
or by D-brane intersections. But besides much progress in
understanding supersymmetric string and gauge theories during the 
past five years, we still have not improved on the outstanding problem
of fixing the string theory dilaton and the moduli fields. In perturbative
supersymmetric string models the potential for the dilaton and the moduli 
fields is simply flat. In the recently constructed nonsupersymmetric 
models (see for instance \cite{ads}), even though the flatness of the 
potential is no longer protected by supersymmetry and the moduli
fields may be stabilized, the dilaton is still expected to get a 
simple runaway behaviour.
 
 The racetrack scenario proposed in the 1980's 
\cite{drsw,krasnikov,alberto}, based on gaugino
condensation for several hidden sector gauge groups, remains as the
best option so far for generating a potential for the dilaton field,
giving rise to a nontrivial minimum at weak coupling after fine tuning
the ranks of the different gauge groups. 
In this scenario the dilaton field is stabilized at a supersymmetric
point and the breaking of supersymmetry is usually
achieved by the moduli fields. A detailed study of this scenario was
performed in \cite{alberto}, for a recent discussion see \cite{dine}.

In this note, we will investigate an alternative to the racetrack
scenario based on recent developments in understanding ${\cal N}=1$
supersymmetric theories. In particular, we will concentrate on the
so-called ${\cal N}=1^*$ theory which is obtained from ${\cal N}=4$ 
super Yang-Mills (SYM) after deforming it by the addition of  
nonvanishing mass terms for the three adjoint chiral
superfields inside the ${\cal N}=4$ vector multiplet. 
Due to its close relation to $\cn=4$, it has been possible to 
understand the general phase
structure of this theory which has proven to be very rich, including
all different phases that 't Hooft predicted for a general class of gauge
theories in the past. A general superpotential that
describes the different phases of the theory and transforms
in a well defined way under the  $SL(2,\IZ)$ of the original $\cn=4$
theory was found in \cite{nick}.
It was shown to be the exact superpotential of the effective
low-energy description where all massive degrees of freedom are
integrated out. 
Furthermore, in \cite{ps} a very detailed analysis of this
theory provided  nontrivial evidence for the extension of the AdS/CFT
correspondence to the $\cn=1$ case and uncovered the detailed brane 
realization of the different confining and Higgs phases. The 
nonsingular supergravity dual allows for physical quantities, 
such as different condensates, to be calculable.
The explicit form of the superpotential of \cite{nick}, which includes
an infinite sum over instantons and fractional instantons, also  provides
successful comparisons between the string and field theory pictures \cite{adk}.

Using the information we have about this special $\cn=1$ theory,
and in particular the expression for the exact superpotential for 
each massive phase,
$W(\tau)$ with $\tau=\theta/2\pi+i4\pi/g^2$, 
we may immediately think about the possibility 
that $\tau$ could actually be promoted to a field as it happens in
string theory, where, at tree-level, it corresponds to the dilaton
field.
 This may happen if the $\cn=1^*$ theory could be part of a
string hidden sector and therefore the corresponding superpotential
can be seen as a dynamically generated superpotential for the dilaton
field. The most natural realization of this idea is the world-volume 
gauge theory on a set of D3-branes in type IIB string theory. However,
one may also imagine some heterotic string vacuum having an $\cn=1^*$ 
subsector. Of course, this approach is only self-consistent if the
masses of the adjoint fields of the $\cn=1^*$ theory are by at least
one order of magnitude smaller than the string scale. Else one would
have to take into account the massive string states as well and would
therefore loose the justification to study the $\cn=1^*$
theory. Throughout this article, we assume that such a self-consistent
mass-deformation of $\cn=4$ SYM does exist. At the end of section 3,
we suggest a way how to realize this assumption in a concrete
model. However, in this paper we do not attempt to 
present a full model. Our work rather seeks to
explore the consequences of recently studied
non-perturbative ${\cal N}=1$ dynamics for
the dilaton behavior. Indeed, in this note we
study the dilaton potential of ${\cal N}=1^*$.
  More precisely, we analyse the scalar potential of the
supergravity Lagrangian derived from the superpotential for $\tau$.
The main result is that there are minima that stabilize $\tau$ at weak
coupling.

\section{Generalities}

We will start by briefly recalling the status of the dilaton potential
in string theory.
Let us, for simplicity, concentrate only on the dilaton field.
 In compactified $\cn=1$ supersymmetric models, after a
duality transformation this field corresponds to 
$S=(2\,R_1R_2R_3/(\alpha')^3)e^{-\phi}+ia$, where 
$R_i$ are the compactification radii,
$\phi$ is the original 10D dilaton field and the axion field $a$ is
dual to the original $B_{\mu\nu}$ field. Both $\phi$ and $B_{\mu\nu}$
appear in all string theories, and the scalar potential
for $S$ vanishes to all orders in perturbation theory.

Since the early days of string theory, the stabilization of the
dilaton field has been considered one of the major obstacles preventing
string models from making contact with low-energy physics. Dine and
Seiberg \cite{ds} gave a very general argument in which, whatever 
the source of the potential for $S$ is, it has to be such that it 
runs away to $S\to \infty$. Then, they argued, if there is any 
other minimum at finite $S$ it has to be at strong coupling, 
since $S+S^*\sim 1/g^2$, and therefore, unless there is a 
`natural' fine tuning at work, it is not achievable in weak 
coupling strings.

Over the years there have been several proposals for generating a
potential for the dilaton field 
\cite{drsw,krasnikov,sdual,moore,expensive,dvali,kahler,afiq,as}. 
The most successful so far has been the racetrack scenario where 
it is assumed that the hidden sector of string theory has a 
product group structure. Gaugino condensation for each group 
factor is expected to generate  superpotentials of the form
$W_i=\exp (-6\pi S/b_i)$, with $b_i$ being the one-loop beta function
coefficients of the corresponding gauge theory. Each 
$W_i$ has a clear runaway behaviour but summing the 
different superpotentials $W=\sum{\alpha_i W_i}$ gives rise 
to a scalar potential that may have a nontrivial minimum, 
stabilizing the dilaton. For a hidden group of the form 
$SU(N_1)\times SU(N_2)$, the minimum in global supersymmetry 
occurs at 
\beq
2\pi\,S=\frac{N_1 N_2}{N_1-N_2}\ 
        \log\left(-\frac{\alpha_2N_1}{\alpha_1N_2}\right).
\eeq
The ranks $N_1,N_2$ may be (discretely) fine tuned to get a realistic
value for $S$. It may be seen on general grounds \cite{cvetic,alberto}
that in both global and local supersymmetry, the minimum of the
potential for $S$ is such that supersymmetry is not broken. Then the
breaking of supersymmetry is left to the other fields in the theory,
such as the moduli fields \cite{seminal}. Even with the natural
problems of this scenario, regarding especially the fine tuning
plus other possible cosmological ones \cite{steinhardt,cmp}, this 
is at the moment the best proposal we have for stabilizing the dilaton.

Another proposal that was put forward was the use of $S$-duality.
Assuming that there are self-dual $\cn=1$ models, the possible
induced superpotentials should transform in a definite way under the 
conjectured $SL(2,\IZ)$ transformations and several functional forms
have been considered \cite{sdual,moore}. This proposal has several problems:
first, the fact that the superpotential has to be a modular form of
negative weight implies the existence of poles in the fundamental
domain. If the poles are at $S\to \infty$ \cite{sdual}, then we do 
not recover the weak coupling behaviour expected from the general 
Dine-Seiberg arguments. Otherwise \cite{moore} there are singularities
at finite values of the string coupling without a clear physical
interpretation.
Second, since the self-dual points $S=1,e^{i\pi /6}$ are necessarily
extrema of the invariant potential, then the natural value for $S$ is
of order $S\sim 1$ which means strong coupling. Finally there are no
explicit models which are self-dual under $S$-duality that can provide
the superpotentials used in \cite{sdual,moore}.
   
Most of the work done in the past was based on the heterotic string
but, the dilaton being universal, the situation is similar for 
other string theories. However, it is worth pointing out the 
differences. First, in type I
and type II compactifications, the fundamental scale may be
substantially lower than the Planck scale \cite{witten}. 
Therefore the difference between the supersymmetry breaking 
scale $\Lambda$ and the string scale does not have to be very 
large in order to have a realistic supersymmetry
breaking  at low energies. Second, the expression for the gauge
couplings
may differ from just being $f=S$ at tree level. It has been found in
orientifold models that the gauge couplings at tree level depend not
only on $S$ but also on the blowing-up modes $M$: $f=S+kM$ and,
depending on the dimension of the brane that hosts the gauge group, 
the gauge coupling may not depend on $S$ at all but take the form 
$f=T+kM$ \cite{afiv}. The effective action after gaugino condensation 
will then be much richer than in the pure heterotic case because of this 
dependence \cite{afiq,as}. Also since these models are not self-dual under 
$T$-duality, the threshold corrections to the gauge coupling 
differ from those of the heterotic string.
Finally, because the fundamental scale may be much lower than the
Planck scale, there exist now quasi-realistic string models which are
already nonsupersymmetric, which in principle can lift the flat
directions. This is expected for the moduli fields since there may be
a particular radius which minimizes the energy (see the fourth paper 
of reference \cite{ads}\ for a concrete potential). But since these 
are weak coupling vacua, the dilaton potential is just the runaway 
and nonperturbative effects may still be needed to stabilize the 
dilaton. A complete analysis of all of these situations is beyond 
the scope of this note and we will only concentrate 
on the particular case $f=S$.

\section{The $\cn=1^*$ Theory}

Let us review the main aspects of the $\cn =1^*$ theory. The starting 
point is $\cn=4$, $SU(N)$ super Yang-Mills, which in terms of $\cn=1$ is a 
gauge theory with three massless adjoint chiral multiplets $\Phi_i$
and superpotential
\beq
W\ =\ \epsilon_{ijk}\,\Tr{\Phi_i [\Phi_j,\Phi_k]}.
\eeq
Deforming this theory by nonzero mass terms for the fields $\Phi_i$,
\beq  \label{mass_def}
\Delta W\ = \  m_1\,\Tr\Phi_1^2+\ m_2\,\Tr\Phi_2^2\ + m_3\,\Tr\Phi_3^2,
\eeq
breaks supersymmetry    to $\cn=1$ (unless $m_1=0, m_2=m_3$ which gives 
$\cn=2$). The generic case where none of the masses vanishes is
called $\cn=1^*$ \cite{nick,ps,adk}.

The classical vacua of this theory can be found by solving 
$\partial W/\partial\Phi_i\ =\ 0$, which leads to 
\beq  \label{eom}
[\Phi_i, \Phi_j] = \epsilon_{ijk}\,m_k\,\Phi_k.
\eeq
Therefore  the fields $\Phi_i$ are $N$-dimensional representations of
the $SU(2)$ algebra and there is a vacuum for each representation.
Since there is one irreducible representation $SU(2)$ for every
dimension $d$ ($d=2j+1$), the number of vacua is determined by 
the number of partitions of $N$. The irreducible $d=N$ 
representation breaks the $SU(N)$ symmetry completely and it is 
identified then with the Higgs phase.
The identity representation ($\vev{\Phi_i}=0$) leaves the 
full gauge group unbroken, 
corresponding to the confining phase of the theory.

All other $N$-dimensional reducible representations represent 
intermediate cases. For instance, if we have a product of two 
irreps of dimension $p_1$ and $p_2$ (with $p_1+p_2=N$), the 
generator $-p_2\id_{p_1}\oplus p_1\id_{p_2}$ is left invariant,
generating a $U(1)$ symmetry. If there are $l$ of these blocks, there will
be a remaining $U(1)^{l-1}$ symmetry left. These then correspond to
the Coulomb phases of the theory. However, if above we have 
$p_1=p_2=N/2$, there will be two extra off-diagonal generators 
left invariant promoting the symmetry to $SU(2)$ and in the general 
case of $q$ blocks of dimension $p$ with $pq=N$, 
this generalizes to a nonabelian $SU(q)$ symmetry left 
invariant. These are the confining phases of the corresponding
$SU(q)$ remnant theory which will have a mass gap and can then be
called the massive phases to differentiate them from the Coulomb phases
which have no gap. The massive phases are then labelled by the two
integers
$p,q$ (with the original confining phase 
corresponding  to $q=N$). Since $pq=N$, the phases are determined by
the divisors of $N$. Furthermore, Donagi and Witten
\cite{dw} found that the quantum vacua are such
that for each $q$, the $SU(q)$ theory in turn splits into $q$
different vacua labelled by an integer $k=0, \cdots , q-1$. These $q$ vacua,
unlike the case of pure super Yang-Mills, are not
equivalent. Therefore we label the massive phases as $(p,q,k)$.

The massive phase structure of the $\cn=1^*$ theory has been shown
\cite{dw} to be rich enough as to realize all the different phases of 
a class of 
gauge theories classified by 't Hooft in the past \cite{thooft}. He 
found that the vacua of an $SU(N)$ gauge theory in which all fields
transform trivially under the centre of $SU(N)$ are in one-to-one
correspondence with the order-$N$ subgroups of
$\IZ_N\times\IZ_N$. Using the integer parameters $p,q,k$, introduced
above, this correspondence can be made precise in the $\cn=1^*$ theory
\cite{dw}: The phase labelled by $(p,q,k)$ is associated with the
$\IZ_N\times\IZ_N$ subgroup generated by $(p,k)$ and $(0,q)$.
The $SL(2,\IZ)$ $S$-duality of the original $\cn=4$ theory,
$\tau\rightarrow (a\tau+b)/(c \tau +d)$, $ad-bc=1$,  is no longer
a symmetry of the $\cn=1^*$ theory but still acts in a
very interesting way on these phases, permuting the phases among
themselves. Under an $SL(2,\IZ)$ transformation, the phase $(p,q,k)$ 
is mapped to the phase $(p',q',k')$, where $(p',k')$, $(0,q')$ are the
(standard) generators of the order-$N$ subgroup that is also generated
by 
\beq \label{SL2Z_action}
\left(\tilde k\atop \tilde p\right)=\left(\begin{array}{cc} a&b\\ 
                                                c&d\end{array}\right)
\left(k\atop p\right),\qquad
\left(\tilde q\atop \tilde r\right)=\left(\begin{array}{cc} a&b\\ 
                                                c&d\end{array}\right)
\left(q\atop 0\right).
\eeq
It is straightforward to show that this implies
\beqa \label{phasemap}
\tau\to \tau+1 &\quad{\rm maps}\quad 
               &(p,q,k)\to(p,q,k+p), \nonumber\\
\tau\to-{1\over\tau} &\quad{\rm maps}\quad 
                     &(p,q,k)\to(\alpha,{N\over\alpha},k'),\\
&&{\rm with}\ \alpha=gcd(k,q). \nonumber
\eeqa
In general, the dependence of $k'$ on $p,q,k$ is complicated, but 
there are two simple special cases: $k'=0$ if $k=0$ and $k'=N-p$ 
if $k=1$. The first statement implies that the transformation 
$\tau\to-1/\tau$ exchanges Higgs and confinement phases.

In \cite{nick} an exact superpotential was derived for this theory
by using instanton techniques for the theory compactified on a 
circle. The compactification to three dimensions is a computational 
trick and it turns out that the superpotential is independent of the 
compactification radius. After integrating out the gauge fields,
it takes the form\footnote{We suppressed a dimensionful constant
overall factor $(N^3/24)m_1m_2m_3$, where $m_i$ are the masses
of the three adjoint chiral superfields \cite{adk}. Also, an
additive holomorphic contribution $A(\tau,N)$ is in principle
possible \cite{adk}. This contribution spoils the modular 
properties of $W$ and will not be considered in the present 
note.}:
\beq
\label{master}
W_{p,q,k}(\tau)= E_2(\tau) - \frac{p}{q}\
E_2\left(\frac{p}{q}\tau+\frac{k}{q}\right)
\eeq
Here $E_2$ is the holomorphic second Eisenstein series:
\beq
E_2(\tau)\ =\ {3\over\pi^2}\ \sum_{n=-\infty }^{\infty } \ 
              \sum_{m=-\infty }^{\infty }\ \frac{1}{(m+n\tau)^2},
\eeq
where the sum excludes the term $m=n=0$.
As it is clear, each phase of the theory has a different
superpotential and even though each $E_2$ series in (\ref{master}) 
does not transform covariantly under $SL(2,\IZ)$ modular
transformations, their difference is a modular form of weight two 
up to some permutation of the phases. 
Notice that $SL(2,\IZ)$ is $not$ a symmetry of the theory but it maps
different phases to one another. The superpotential
reflects the $SL(2,\IZ)$ properties of the theory in an interesting
way. For instance, it is clear that under $\tau\rightarrow \tau+1$,
$W_{p,q,k}\to W_{p,q,k+p}$ and, more generally, one can show that
\beq \label{Wtransform}
W_{p,q,k}\left({a\tau+b\over c\tau+d}\right)=
(c\tau+d)^2 W_{p',q',k'}(\tau),
\eeq
where $p',q',k'$ are determined as explained in the paragraph 
before eq.\ (\ref{SL2Z_action}).

Before discussing the minimization of the scalar potential in detail
in the following section, we would like to give a qualitative argument 
why we expect the scalar potential to have a minimum at large values 
of $\tau$. The Eisenstein series has a rich structure at values of its
argument of order $1/2\pi$ and it drops exponentially to a constant for 
large values of its argument. Thus, for $p\ll q$, the first $E_2$ in 
(\ref{master}) can be approximated by a constant and the second $E_2$ 
has a rich structure at $\tau\gg1/2\pi$. Besides the constant term in
the expansion, $W$ then looks like an infinite sum of gaugino condensates.

To finish this section, let us address the following issue: as
mentioned in the introduction we are assuming that the masses of the
adjoint scalar fields are smaller than the string scale in
order to be able to consider the $\cn=1^*$ as an effective field
theory below the string scale. Let us then 
comment on how the $\cn=1^*$ theory
appears in string theory and why the masses of the adjoint 
fields do not necessarily have to be of the order
of the string scale.
 The most natural realization is on a set of
$N$ D3-branes. We think of these D3-branes as filling the 3+1
space-time dimensions and being located at a nonsingular point of some
6-dimensional compact space. The theory living on the D3-branes is an
$\cn=4$ $U(N)$ gauge theory. We know of two ways to switch on a mass
deformation that breaks the supersymmetry down to $\cn=1$.

First, we will  consider the $\cn=1^*$ theory as it arises in the
context of the AdS/CFT correspondence.
As discovered by Myers \cite{myers}, D$p$-branes can couple
to $(p+3)$-form RR potentials. In particular, consider $N$ D0-branes
in flat space. The equations of motion of the 9 adjoint fields
$\phi_i$, $i=1,\ldots,9$, whose expectation values parametrise the
positions of the D0-branes are $[\phi_i,\phi_j]=0$ $\forall\,i,j$. All
the $\phi_i$ can be simultaneously diagonalized and the $N$
eigenvalues of $\vev{\phi_i}$ are interpreted as the $i^{\rm th}$
coordinate of the positions of the D0-branes. Turning on a background
flux of the RR 4-form field strength
\beq \label{fourflux}
   F^{(4)}_{0ijk}=\left\{\begin{array}{ll} -2\,m\,\epsilon_{ijk}
                             &{\rm if}\ i,j,k\in\{1,2,3\}\\
                           0 &{\rm else} \end{array}\right.
\eeq
changes the scalar potential of the gauge theory on the D0 world-line
and leads to the modified equations of motion \cite{myers}
\beq \label{mod_eom}
  [\phi^i,\phi^j]=im\,\epsilon_{ijk}\,\phi^k,\quad{\rm for}\ 
        i,j,k\in\{1,2,3\}.
\eeq
The geometry of this configuration is noncommutative. It describes a
fuzzy 2-sphere of radius
\beq  \label{fuzzy_radius}
  R=\left(\sum_{i=1}^3 \frac{1}{N} \Tr(\phi_i^2)\right)
   =\frac{m}{2}N\sqrt{1-\frac{1}{N^2}}.
\eeq
The equations of motion (\ref{mod_eom}) of the adjoint fields
$\phi_1$, $\phi_2$, $\phi_3$ living on the D0 world-line remind us
the equations of motion (\ref{eom}) of the adjoint fields of
$\cn=1^*$. Indeed, Polchinski and Strassler \cite{ps} showed that the
Myers effect applied to D3-branes in AdS$_5\times S^5$ space-time
leads to the expected mass deformation (\ref{mass_def}) for the
adjoint fields of the $\cn=4$ SYM. More precisely, switching on a 
background flux of the RR 7-form field strength of the form 
$F^{(7)}_{0123ijk}=\alpha\,m_k\,\epsilon_{ijk}$ if $i,j,k\in\{4,5,6\}$
implies (\ref{eom}) for the adjoint fields phase rotated to their real
parts. Here, $\alpha$ contains some numerical factors due to the
AdS$_5\times S^5$ space-time. At this level, the RR 7-form is an
arbitrary parameter which can be chosen such that the masses are
significantly smaller than the string scale. Therefore the RR flux can
be seen as providing an independent scale from the string scale 
in much the same way as the
compactification scale does not have to be the same as the string
scale.

Second, one can foresee the following scenario:
 take the compact six dimensions to form an orbifold
space that preserves $\cn=1$ supersymmetry. Putting $N$ D3-branes in
the bulk, i.e., at some nonsingular point of the orbifold, results in
an $\cn=4$ SYM on their world-volume. Let us denote by $A$ this set of
D-branes and put another set --- denoted $B$ --- at a singular point
of the orbifold. The gauge theory on the latter will only be $\cn=1$
supersymmetric due to the orbifold action.
 The two sectors of the model only interact
gravitationally. As a consequence, the partial supersymmetry breaking
from $\cn=4$ to $\cn=1$ in the $B$-sector will be transmitted to the
$A$-sector via gravitational interactions. This will give masses to
the adjoint chiral superfields $\Phi_i$. Using dimensional analysis,
the size of the masses can be estimated to be of the order $m\sim
M_{\rm str}^2/M_{\rm Pl}$, where we used that supersymmetry is
partially broken at the string scale in the $B$-sector. Thus, for low
or intermediate string scale, the masses of the adjoint fields may be 
naturally
suppressed with respect to the string scale. As we mentioned in the
introduction, the explicit construction of string models with these
characteristics is beyond the scope of the present work.

\section{The Dilaton Potential}

We will now promote the parameter $\tau$ to a full $\cn=1$ superfield
that as mentioned before may be several combinations of the dilaton and 
other moduli fields in different string theories. In this note we focus
on the dilaton and set $S=-i\tau=(2\,R_1R_2R_3/(\alpha')^3)e^{-\phi}+ia$.
We consider the superpotential
\beq
W_{p,q,k}(S)=E_2(S) - \mu E_2(\tilde S),
\label{wdef}
\eeq
where $\tilde S = \mu S - i\nu$, $\mu=p/q$ and $\nu=k/q$.
In the following we will usually drop the sub-indices in $W$.
We will mostly work with the $E_2(S)$ expansions in terms of the
variable $\exp(-2\pi\,S)$ that are given in the appendix. Clearly
$W(S)$ is periodic in ${\rm Im}\, S$ with a period that depends on the
particular values of $p,q$. Notice that $W$ goes to
the constant $(1-\mu)$ for large $S$. 
On the other hand, $W$ diverges as $1/S^2$ for small $S$,
as found using (\ref{e2mod}).

Our purpose is to study the scalar potential $V(S)$ which also depends
on the K\"ahler potential that determines the dilaton kinetic energy.
We take the weak coupling result in 4-dimensional string models, namely
\beq
K=-\log (S+S^*), 
\label{kpdef}
\eeq
and neglect possible perturbative and nonperturbative corrections.

\begin{figure}
\begin{center}
\centering
\epsfxsize=12cm
\epsfysize=9cm
\leavevmode
\epsfbox{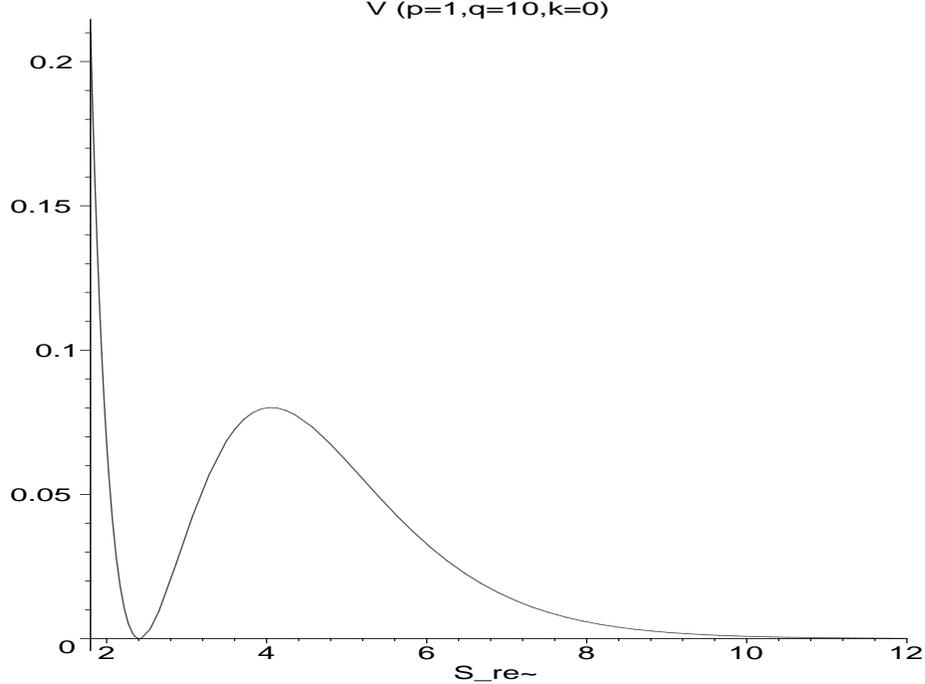}
\end{center}
\caption[]{Dilaton potential in global supersymmetry.}
\label{globv}
\end{figure}

Let us first discuss the case of global supersymmetry in which
\beq
V = K^{-1}_{SS^*}|W_S|^2 = (S+S^*)^2 |W_S|^2 
\label{vglob}
\eeq
Using the formulae provided in the appendix we find
\beq
W_S =  \frac{\pi}6 \left\{ [E_4(S) - E_2^2(S)] - \mu^2
[E_4(\tilde S) - E_2^2(\tilde S)] \right\}. \label{wder}
\eeq
For large $S$, $W_S$ tends to zero as $\exp(-2\pi\,S)$, whereas for
small $S$ it diverges as $1/S^3$. Hence, $V$ diverges at small
$S$ and vanishes at large $S$. Our numerical analysis indicates
that in between $V$ has a minimum and its behaviour in the 
${\rm Re}\, S$ direction is of the form
shown in figure \ref{globv}. At the minimum supersymmetry is not broken
since $\langle W_S \rangle=0$. For the class of models with $p=1$ and
$q=N$ we find that the supersymmetric minimum is located at
\beq
{\rm Im}\, S=\frac{N}2 + k, \quad\quad\quad\quad
{\rm Re}\, S\approx 0.24104\,N,
\label{globmin}
\eeq
which corresponds to weak coupling for $N\ge5$. This result can be 
derived semi-analytically by noticing that for large $S$ the second 
term in $W$ dominates. Therefore the supersymmetric minimum is to very good
approximation located at values $S_{\rm min}$ that satisfy 
$E_2'({S_{\rm min}-ik\over N})=0$, where the prime denotes derivative.
The numerical analysis leads us to believe that there is only
one minimum (up to the periodicity $S\to S+iN$), but we were not
able to prove this.

Let us now turn to the case of local supersymmetry. The scalar potential
and its first derivative turn out to be
\beqa
V & = & \frac1{S_R} \left\{ |S_R W_S  - W|^2 - 3|W|^2 \right \}, 
                                                     \nonumber \\[0.1cm]
\frac{\partial V}{\partial S} & = & W_{SS} (S_R W_S - W)^* - 
                 \frac{2W^*}{S_R^2} (S_R W_S - W),
\label{vloc}
\eeqa
where $S_R=S+S^*$. Notice then that there
can be two types of extrema. The supersymmetric extrema appear when
\beq
S_R W_S - W = 0.
\label{msloc}
\eeq
These extrema are minima
provided that $S_R^2|W_{SS}| > 2 |W|$.
The nonsupersymmetric extrema occur at
\beq 
S_R^2 W_{SS} = 2 W^* e^{2i\gamma}, 
\label{mnsloc}
\eeq
where $\gamma=\arg(S_RW_S - W)$. As discussed below, our numerical analysis 
indicates that all minima satisfy (\ref{msloc}) so that they are supersymmetric
and lead to negative cosmological constant.

We have explored to some detail the class of models with $p=1$ and $q=N$
for which it is enough to set $k=0$ since $k\not=0$ can be reached
by a translation $S \to S-ik$. In order to perform reliable computations 
with the Eisenstein series we use the weak coupling expansion of $W(S)$
in eq. (\ref{wdef}) only when $2\pi\,{\rm Re}\, S > N$. For other
ranges we can use the property (\ref{e2mod}). For instance, for
$1 < 2\pi\,{\rm Re}\, S < N$ we can transform the $E_2(\frac{S}{N})$
term to obtain
\beq
W(S)=E_2(S) + \frac{N}{S^2} E_2(\frac{N}{S}) - \frac{6}{\pi S}.
\label{w2def}
\eeq
Similarly, when $2\pi\,{\rm Re}\, S < 1$ we can transform both terms
in $W(S)$ to obtain
\beq
W(S)=-\frac{1}{S^2}E_2(S) + \frac{N}{S^2} E_2(\frac{N}{S}).
\label{w1def}
\eeq 
For $N \leq 8$ we find one supersymmetric minimum
at ${\rm Im}\, S=\frac{N}2$ and ${\rm Re}\, S$ given in table~\ref{ta1}.
\begin{table}[htb]
\begin{center}
\begin{tabular}{|c|ccccccc|}
\hline
$N$ & 2 & 3 & 4 & 5 & 6 & 7 & 8 \\
\hline
${\rm Re}\, S$ &  1.81 & 2.29 & 2.70 & 3.04 & 3.30 & 3.45 & 3.37\\
\hline
\end{tabular}
\end{center}
\caption{Minima of the scalar potential.}
\label{ta1}
\end{table}
For $N \geq 9$ this minimum turns into a saddle point and two
supersymmetric minima $S_{11}$, $S_{12}$ on either side in the 
${\rm Im}\, S$ direction appear, such that 
${\rm Im}\, S_{12} + {\rm Im}\, S_{11} = N$ and 
${\rm Re}\, S_{12} ={\rm Re}\, S_{11}$.
Examples are given in tables~\ref{ta2}, \ref{ta2b}.
\begin{table}[htb]
\begin{center}
\begin{tabular}{|c|cccccccccc|}
\hline
$N$ & 10 & 20 & 30 & 40 & 50 & 60 & 70 & 80 & 90 & 100 \\
\hline
${\rm Re}\, S_{11}$ & 3.08 & 3.95 & 4.50 & 4.92 & 5.30 & 5.65 & 5.96 
                    & 6.26 & 6.53 & 6.79\\
${\rm Im}\, S_{11}$ & 4.07 & 6.11 & 7.49 & 8.66 & 9.69 & 10.61 & 11.45
                    & 12.23 & 12.97 & 13.66 \\
\hline
\end{tabular}
\end{center} 
\caption{Minima of the scalar potential.}
\label{ta2} 
\end{table} 

\begin{table}[htb]
\begin{center}
\begin{tabular}{|c|cccccccccc|}
\hline
$N$ & 200 & 300 & 400 & 500 & 600 & 700 & 800 & 900 & 1000 &2000 \\
\hline
${\rm Re}\, S_{11}$ &  8.89 & 10.49 & 11.84 & 13.03 & 14.09 & 15.09 
                    & 16.01 & 16.86 & 17.67 & 24.23\\
${\rm Im}\, S_{11}$ & 19.21 & 23.45 & 27.02 & 30.16 & 32.99 & 35.60 
                    & 38.05 & 40.30 & 42.45 & 59.81 \\
\hline
\end{tabular}
\end{center} 
\caption{Minima of the scalar potential.}
\label{ta2b} 
\end{table} 
For large $N$ (say $N\ge20$) these results agree well with those 
obtained using for $W$ the approximation
\beq
W \sim 1 + \frac{N}{S^2} - \frac{6}{\pi S}.
\label{waprox}
\eeq
This approximation is derived from (\ref{w2def}) by keeping only
the constant terms in the $E_2$ expansions. It is valid when
$(2\pi\,{\rm Re}\, S)\,N\gg({\rm Re}\, S)^2+({\rm Im}\, S)^2$.

Now, as $N$ grows, there appear more supersymmetric minima, always in
pairs $S_{i1}$, $S_{i2}$ with ${\rm Im}\, S_{i2} + {\rm Im}\, S_{i1} = N$ and
${\rm Re}\, S_{i2} ={\rm Re}\, S_{i1}$, $i=1,\cdots, n$. We have not
performed an extensive numerical analysis and thus limit ourselves to 
pointing some observations. The number $n$ of minima grows with $N$.
For instance, we found $n=8$ for $N=100$, although we cannot exclude 
that there exist even more minima. 
It turns out that the absolute minimum is among the new minima that have
${\rm Im}\, S_{i1} \geq {\rm Im}\, S_{11}$ and
${\rm Re}\, S_{i1} \leq {\rm Re}\, S_{11}$. We label by $S_{n1}$ the
minimum with the lowest value of the potential that we found. 
In table ~\ref{ta3} we give the values of these minima in various cases.
\begin{table}[htb]
\begin{center}
\begin{tabular}{|c|ccccccccc|}
\hline
$N$ & 20 & 30 & 40 & 50 & 60 & 70 & 80 & 90 & 100 \\
\hline
${\rm Re}\, S_{n1}$ & 2.38 & 3.05 & 2.41 & 2.76 & 3.04 & 2.49 & 2.69 & 2.88 & 3.04\\
${\rm Im}\, S_{n1}$ & 7.46 & 11.66 & 11.07 & 13.99 & 16.91 & 15.29 & 17.54 & 19.80 &
22.05 \\
\hline
\end{tabular}
\end{center}
\caption{Further Minima of the scalar potential.}
\label{ta3}
\end{table}

Using the approximate superpotential in (\ref{waprox}), the condition
for a susy minimum, {\it i.e.},\ $D_SW=0$, reduces to a cubic equation 
which can be solved analytically. The general solution is very 
complicated and not illuminating but there is a nice expansion 
in powers of $\sqrt N$. One has\footnote{Only one of the three 
solutions of the cubic equation gives a positive 
${\rm Re}\, S_{\rm min}$ in the range of validity of the
approximation.}
\beq \label{Smin}
    2 \pi\, {\rm Re}\, S_{\rm min} = \sum_{n=1}^\infty c_n 
                           \left(\pi \sqrt N\right)^{2-n},
\eeq
where the first coefficients are 
 $ c_1=1, c_2=12, c_3=-36, c_4=648$.
Moreover we found a recursion formula for the $c_n$. Given $c_1$ and
$c_2$, they can be obtained from 
\beq
     c_r = -\frac12 \sum_{\scriptstyle n+m+k=r+2\atop\scriptstyle  n,m,k<r} 
                       \!\!\!\!\!\!c_n c_m c_k
             +9 \sum_{n+m=r} \!\!c_n c_m
            -36\, c_{r-2}.
\eeq
The solution for the imaginary part of $S$ is similar:
\beq
 2 \pi\, {\rm Im}\, S = \frac{1}{\sqrt7}\  \sum_{n=1}^\infty d_n 
                         (7 \pi \sqrt N)^{2-n},
\eeq
with 
$   d_1=1, d_2=12, d_3=-1332, d_4=90072, \ldots$

If one neglects the term $6/(\pi S)$ in (\ref{waprox}), the
condition for a susy minimum reduces to a quadratic equation with a very
simple solution:
    $ {\rm Re}\, S = 1/2 \sqrt{N}, \ 
     {\rm Im}\,  S = 1/2 \sqrt{7N}$.
This is just the first term in the above expansions, but the approximation
is rather rough. The series converges very slowly. But in general it
will
give us an idea for the values of $N$ required to get `realistic'
gauge couplings. In the standard unification picture we need 
${\rm Re}\,S\sim 20$ indicating that a very large value of $N$ may be
needed. This is not possible to get in
perturbative heterotic strings but in more general vacua including
$F$-theory and orientifold models this is easily achievable \cite{kl}. 
We present in the figures  examples of the general behaviour of
the potentials regarding periodicity, runaway to   infinity and 
supersymmetric minima at weak coupling. We explicitly show the case 
$q=N=1000$ for different ranges of values of Re$\,S$ and Im$\,S$
around the supersymmetric weak coupling minimum.
\begin{figure}
\begin{center}
\centering
\epsfxsize=10cm
\epsfysize=5cm
\leavevmode
\epsfbox{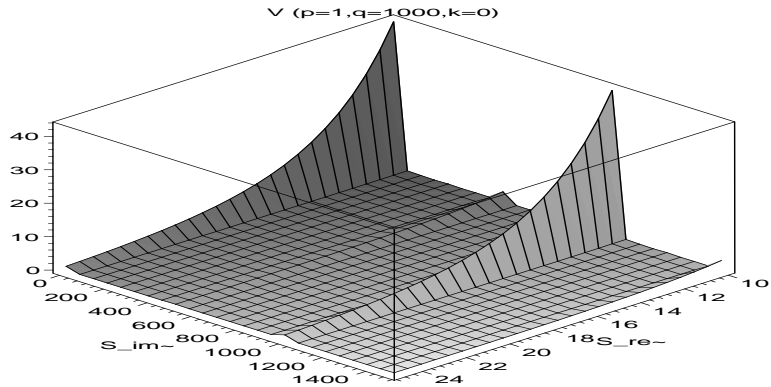}
\end{center}
\caption[]{\small Case $q=N=1000$: Periodicity of the  potential.}
\label{locv1}
\end{figure}

\begin{figure}
\begin{center}
\centering
\epsfxsize=10cm
\epsfysize=5cm
\leavevmode
\epsfbox{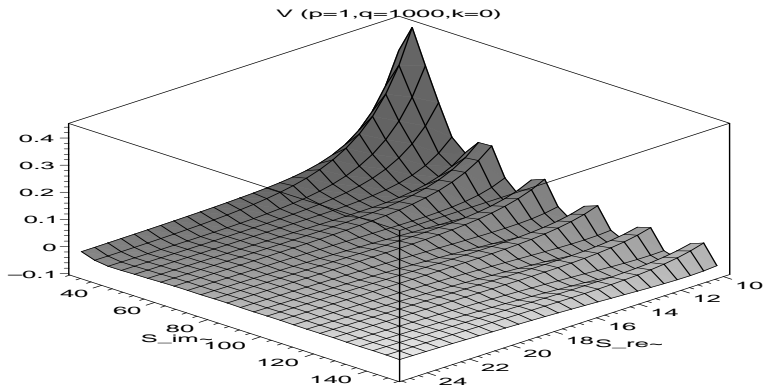}
\end{center}
\caption[]{\small Case $q=N=1000$: Structure of the potential.}
\label{locv2}
\end{figure}

\begin{figure}
\begin{center}
\centering
\epsfxsize=10cm
\epsfysize=5cm
\leavevmode
\epsfbox{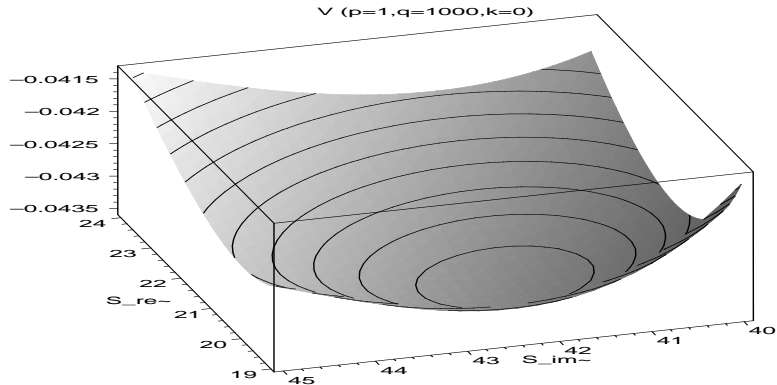}
\end{center}
\caption[]{\small  Case $q=N=1000$: Minimum of the  potential.}
\label{locv3}
\end{figure}

The case $p=N$, $q=1$ is easily derived from the above. Indeed,
notice that $W_{N,1,0}(S)=-N\,W_{1,N,0}(NS)$ and hence the minima
will be those found previously divided by $N$.
We have also considered a few examples with $N\not= p \not=1$ and 
found similar results. For example, for $p=2$, $q=4$, there is a 
supersymmetric minimum at ${\rm Re}\, S=1.8103$ and ${\rm Im}\, S=1$, 
whereas for for $p=2$, $q=5$, there are two supersymmetric
minima at ${\rm Re}\, S_1= {\rm Re}\, S_2=2.0562$ and 
${\rm Im}\, S_1=5 - {\rm Im}\, S_2=1.2489$.
Finally the, case $p=q$ and $k \not=0$ has also supersymmetric minima
that generally occur at small ${\rm Re}\, S$ and lead to larger 
values of the cosmological constant.

Regarding nonsupersymmetric extrema, all the ones we have found
correspond to maxima or saddle points. So it seems that all the vacua
of this theory are supersymmetric,  but we still lack a proof that 
this is the general situation.

\section{Discussion}

We have seen that the nonperturbative dilaton potential 
for $\cn=1^*$ theory has very interesting properties.
In particular, it can stabilize the dilaton field at weak string
coupling in a  natural way. The discrete fine tuning required to
overcome the Dine-Seiberg problem is naturally provided in the theory
by the integers $p,q,k$ determining the phases. We have seen that the
minima we found generically do not break supersymmetry, which  is 
similar to the situation of the racetrack scenario. This situation
may change once corrections to the K\"ahler potential are included,
perturbative and nonperturbative. Furthermore, once the other moduli
fields $T$ are included, they tend to break supersymmetry. Our results for
the minima in the $S$ direction will be preserved whenever the
superpotential is a product $W(S,T)=\Omega(S)\Gamma(T)$
and the K\"ahler potential does not have a large  $S-T$ mixing.
 This is certainly the case in
perturbative heterotic strings where $\Gamma(T)$ comes from threshold
corrections to the gauge coupling constant. In the simplest case of
constant $\Gamma(T)$ we have a realization of the no-scale scenario with
supersymmetry broken and vanishing cosmological constant and $T$
arbitrary.
More generally, $T$ will be fixed with a negative cosmological
constant \cite{seminal}.

Notice that the situation we have has some similarities with the
racetrack scenario, although perhaps improving on the discrete fine
tuning part. It shares some of the positive properties, such as the
overcoming of Dine-Seiberg problem and fixing the dilaton without
breaking supersymmetry. It may also share some of the problems.
In particular, the cosmological constant does not have to be
small at the minimum after supersymmetry is broken. Furthermore the
cosmological problems emphasized in \cite{steinhardt}\  about the dilaton
field overrunning the nontrivial minimum towards the runaway one for
generic initial conditions, may still hold  in
this case as well as the proposed ameliorations \cite{barreiro},
similarly for the cosmological moduli problem \cite{cmp}.

 There are also some significant differences, besides the theoretical
one of having an underlying $SL(2,\IZ)$ symmetry mapping the different
massive phases. In particular the constant term
in the superpotential is not present in the many condensates case. It is
actually more reminiscent of the original stringy gaugino condensation
discussions with a constant term added from the antisymmetric tensor
field \cite{drsw}.
 The presence of this constant term has at least one important impact:
after supersymmetry is broken, the scale of supersymmetry breaking
is not exponentially supressed as compared to the string scale as in
gaugino condensation models where $M_{SUSY}=M_{String}e^{-b/g^2}$. In
our case it will depend on how supersymmetry is broken. But if
the scale is proportional to the superpotential, the constant term
will dominate over the exponentially suppressed ones and both scales
will be similar. This may be consistent with a low or intermediate 
scale for string theory \cite{witten,biq}.

Probably the clearest difference from previous proposals is the large
proliferation of supersymmetric vacua at large $N$ on top of the
standard repetition of minima from the periodicity of the potential.
This may have  some interesting consequences, especially regarding 
cosmology. The many vacua structure is such that we have a realization
of a staircase potential, once we project on some line in the
Re$\,S$\,/\,Im$\,S$ plane, with the staircase climbing towards the 
zero coupling vacuum. Each vacuum 
has different (negative) energy but all of them are stable according 
to the general analysis of Coleman-de Luccia \cite{weinberg}.
We can imagine then many different regions of the universe living on
different vacua with static BPS domain wall boundaries. 
The critical tension of the domain walls is determined by the 
difference of the value of the superpotentials for
the corresponding minima \cite{cqr}.
 In general, the vacua are stable if the tension of the
wall is bounded by the BPS condition, but the walls are not static
if the BPS condition is not satisfied. The general properties of these
domain walls have been investigated thoroughly (see \cite{mirjam} for a
review on domain walls in supergravity theories).
  For some interesting recent ideas along these lines see
\cite{banks}. 

Once, more fields are considered, the physical implications of the
 proliferation of vacua changes. In the simple no-scale models for
 instance, all the vacua would be degenerate with vanishing
 cosmological constant and broken supersymmetry. We may even foresee
that the addition of other fields and corrections to the K\"ahler
 potential may revert the staircase behaviour producing a potential 
similar to the one proposed by Abbott \cite{abbott}
 trying to ameliorate the
 cosmological constant problem (this may happen for instance if there
 is a field $U$ not entering in the superpotential
with K\"ahler potential $K=-n \log(U+U^*)$ and $n>3$, although we do
 not see how this situation may be realized in string theory).

An interesting open question would be to find an explicit realization
of the present scenario in a concrete string model by looking for hidden
sectors, perhaps on a set of hidden D3-branes corresponding to 
$\cn=1^*$. One might hope however that since
the massive phase structure of the $\cn=1^*$ theory is very rich and realizes
all the phases predicted by 't Hooft, the results we have found in
this note may also apply to more general $\cn=1$ theories.

\vskip1cm
\centerline{\bf Acknowledgements}

We thank Ralph Blumenhagen, Cliff Burgess, Elena C\'aceres, Massimo Bianchi, 
Nick Dorey, Michael Green, Luis Ib\'a\~nez, Stefano Kovacs, Prem Kumar,
Dieter L\"ust, Graham Ross and Radu Tatar
for useful conversations. This research is partially funded by PPARC.
A.F. acknowledges a fellowship from the Alexander von Humboldt Foundation. 

\vskip1cm

\section{Appendix}

In this appendix we collect some definitions and useful properties.

The Eisenstein modular functions that enter in the scalar potential
and its derivatives are
\beqa
E_2 & = & 1 - 24 \sum_{n=1}^\infty \sigma_1(n) q^n, \nonumber \\[0.1cm]
E_4 & = & 1 + 240 \sum_{n=1}^\infty \sigma_3(n) q^n, \nonumber \\[0.1cm]
E_6 & = & 1 - 504 \sum_{n=1}^\infty \sigma_5(n) q^n,
\label{efns}
\eeqa
where $q=e^{-2\pi\,S}$ and $\sigma_p(n)$ is the sum of the $p^{\rm th}$
powers of all divisors of $n$. $E_4$ and $E_6$ are modular forms of 
weight 4 and 6 respectively. This means
\beq
E_4\left(\frac1{S}\right) = S^4 E_4(S), \quad\quad\quad\quad
E_6\left(\frac1{S}\right) = -S^6 E_6(S).
\label{emod}
\eeq
$E_4(S)$ has a zero at $S=e^{i\pi/6}$ and $E_6(S)$ at $S=1$.
$E_2$ fails to be a modular form of weight two since
\beq
E_2\left(\frac1{S}\right) = -S^2 E_2(S) + \frac{6S}{\pi}.
\label{e2mod}
\eeq
It is useful to introduce
\beq
\hat E_2(S) = E_2(S) - \frac6{\pi(S+S^*)}, \quad\quad\quad\quad
\hat E_2\left(\frac1{S}\right) = -S^2 \hat E_2(S). 
\label{he2} 
\eeq
$\hat E_2(S)$ has zeroes at $S=1, e^{i\pi/6}$.

We also have the following derivatives 
\beqa
E_4^\prime & = & \frac{2\pi}3 (E_6 - E_2 E_4), \nonumber \\[0.1cm]
E_2^\prime & = & \frac{\pi}6 (E_4 - E_2^2), \nonumber \\[0.1cm]
E_2^{\prime\prime} & = & \frac{\pi^2}{18} (2E_6 - 3E_2 E_4 + E_2^3),
\label{eder}
\eeqa
where prime is $\partial/\partial S$.

\end{document}